\RequirePackage{ifpdf}
\documentclass[12pt,letterpaper]{article}
\pdfoutput=1
\usepackage{jheppub}
\usepackage{epsfig}
\usepackage{enumitem}
\usepackage[latin1]{inputenc}
\usepackage{bbm,amsfonts}
\usepackage{graphicx}
\usepackage{amssymb,amsmath,amsfonts,mathtools}
\usepackage{fancybox}
\usepackage{enumerate}
\usepackage{dsfont}
\usepackage{verbatim}
\usepackage{wrapfig}
\usepackage{slashed}
\usepackage{shuffle}
\usepackage{subcaption}
\usepackage{braket}

\author[a]{Marco S. Bianchi}
 
\affiliation[a]{Niels Bohr Institute, University of Copenhagen, Blegdamsvej 17, 2100 Copenhagen $\emptyset$,\\
Denmark}

\emailAdd{marco.bianchi@nbi.ku.dk}  
  
\abstract{Given the recent progress in computing three-point functions in ${\cal N}=4$ SYM via integrability, I provide here a novel direct calculation of some structure constants at weak coupling.
The main focus is on correlators involving more than one unprotected operator, at two-loop order in the perturbative expansion.}

\title{A note on three-point functions of unprotected operators} 

\newcommand{\be}{\begin{equation}}
\newcommand{\ee}{\end{equation}}
\newcommand{\beq}{\begin{equation}}
\newcommand{\eeq}{\end{equation}}
\newcommand{\bea}{\begin{eqnarray}}
\newcommand{\eea}{\end{eqnarray}}
\newcommand{\ena}{\end{eqnarray}}

\def\Tr{\textrm{Tr}}

\numberwithin{equation}{section}

\def\clock{{\count0=\time
           \divide\count0 60
           \ifnum\count0<10 0\fi\the\count0
           \multiply\count0 -60 \advance\count0 \time
           :\ifnum\count0<10 0\fi \the\count0
         }}
\newcommand{\timestamp}{{\small\vbox{\hbox{\tt\jobname.tex}
\hbox{\the\day/\the\month/\the\year, \clock}}}}
\begin{document}

\maketitle
\allowdisplaybreaks

\section{Introduction}

Computing three-point correlation functions is a crucial and usually hard task in conformal field theories.
In ${\cal N}=4$ SYM integrability \cite{Beisert:2010jr} comes to rescue in the form of the hexagon program \cite{Basso:2015zoa}, which has also been advocated to extend to higher-point functions \cite{Fleury:2016ykk,Eden:2016xvg,Basso:2017khq} and to provide a grasp on non-planar effects \cite{Bargheer:2017nne,Eden:2017ozn}.

On the other hand, direct computations of three-point functions from a standard perturbative expansion at weak coupling have been performed in the literature \cite{Beisert:2002bb,Chu:2002pd,Roiban:2004va,Alday:2005nd,Alday:2005kq,Okuyama:2004bd,Georgiou:2008vk,Grossardt:2010xq,Georgiou:2012zj,Plefka:2012rd} and connection to integrability was established \cite{Escobedo:2010xs,Escobedo:2011xw,Gromov:2011jh,Gromov:2012vu,Gromov:2012uv,Vieira:2013wya}.
However such results are mostly limited to one-loop order.
A powerful alternative approach to their computation uses the OPE of higher-point correlators.
Such a technique has proven extremely successful, especially in the case of correlators of protected operators \cite{Eden:1998hh,Eden:1999kh,Eden:2000mv,Bianchi:1999ge,Bianchi:2000hn,Bianchi:2001cm,Dolan:2001tt}, which can be constructed efficiently \cite{Eden:2011we,Eden:2012tu,Chicherin:2015edu,Chicherin:2018avq}. This has allowed to fix structure constants of two BPS and one unprotected operators to vertiginous loop order, providing spectacular tests and tips on the integrability approach \cite{Eden:2012rr,Eden:2015ija,Basso:2015eqa,Eden:2016aqo,Goncalves:2016vir,Basso:2017muf,Georgoudis:2017meq,Chicherin:2018avq}.

In this note, I tackle the problem of determining structure constants directly (namely without relying on OPE's), with the specific aim of providing perturbative data up to two humble loops, but extending the analysis to three-point functions with more than one unprotected operator. 
In particular, I consider correlators among protected operators in the $SU(2)$ sector, the unprotected Konishi scalar $SU(4)$ singlet, and the unprotected $sl(2)$ Konishi twist-2 operators of spin 2.
Taking correlators of two protected operators and one unprotected I reproduce known results, that have already been computed (and outperformed) from different angles, as mentioned above. For simplicity, I consider here only correlators with up to one operator with spin.
Then a three-point function of two unprotected operators is also allowed in this setting, which has never been computed so far (to the best of my knowledge). Fixing its structure constant at two-loop order is the main focus of this note.

One virtue of the present computation consists in involving no assumptions whatsoever and being not based on any conjecture. Hence it can honestly provide experimental data points for prospective checks of other, possibly more effective, techniques.
The realization of the latter correlator that I mentioned is challenging from the perspective of the hexagon program. On the one hand this is good in the sense that I can provide a complementary computation, producing new results. On the other hand this might not be the best experiment for testing the integrability approach, at the moment.
It would also be interesting to derive the result presented here from an OPE expansion, which would likely provide an easier extension to three-point functions with further operators and higher spins.

\section{Definitions}

I work in ${\cal N}=4$ SYM with gauge group $SU(N)$ and coupling $g$. The planar approximation is not assumed, however the results presented here (namely up to second order in perturbation theory) are not sensitive to sub-leading effects in $N$. Hence the 't Hooft coupling constant $\lambda =\frac{g^2 N}{16 \pi ^2}$ will be used ubiquitously.
The difference between $U(N)$ and $SU(N)$ gauge groups is confined to the tree level pre-factors, that will not play any crucial role in the following.

\subsection{Operators}\label{sec:operators}

I consider twist-2 operators consisting of the complex scalars $X_a$ ($a=1,2,3$) of ${\cal N}=4$ SYM.
The reason why the discussion is limited to low twist originates from the technical simplicity of the computation and is tied to the effectiveness of the computational method, that I spell out in section \ref{sec:strategy}.
In particular, I consider the chiral primary BPS operators
\begin{equation}\label{eq:chiral}
O_{ab} \equiv \Tr (X_a X_b) \qquad,\qquad \bar O^{ab} \equiv \Tr (\bar X^a \bar X^b)
\end{equation}
and the non-chiral operators
\begin{equation} 
O_{a}^{\phantom{a}b} \equiv \Tr (X_a \bar X^b) 
\end{equation}
among which summing over the indices to produce an R-symmetry singlet of naive dimension 2, realizes the Konishi scalar operator.
I shall also use protected operators in this family, taking $SU(2)$ sector-like operators with $a\neq b$.

I also consider twist-2 operators with spin of schematic form
\begin{equation} 
O^j_{ab} \equiv \Tr (D^{k}X_a D^{j-k} X_b) + \dots
\end{equation}
obtained by acting on the chiral operator above with covariant derivatives $D$.
The derivatives are contracted in such a way that they are symmetric and traceless, with the ellipsis indicating the combination with other ways of distributing the derivatives.
These operators are usually projected contracting all indices with a light-like vector $z$. The precise weights of the distribution of derivatives are given by the coefficients of Gegenbauer polynomials
\begin{equation}\label{eq:twist2}
\hat {\cal O}^j_{ab} = \sum_{k=0}^j\, a_{jk}^{\frac{d-3}{2}}\, \Tr\left( \hat D^k X_a \hat D^{j-k} X_b \right) \qquad,\qquad \hat D = D_{\mu} z^{\mu}
\end{equation}
where
\begin{equation}
\sum_{k=0}^j\, a_{jk}^{\nu}\, x^k y^{j-k} = (x+y)^j\, C_j^{\frac{d-3}{2}}\left( \frac{x-y}{x+y} \right)
\end{equation}
and $d$ is the space-time dimension.
For simplicity, I shall always select the same flavor for the fields in these operators with spin and define, say, $\hat {\cal O}^j \equiv \hat {\cal O}^j_{11}$, though the calculations presented here carry through in a similar manner for different choices as well.
In practical calculations I shall only consider here the simplest, spin-2, $sl(2)$ Konishi operator, among this family. For spin $j=0$ the operators are protected and coincide with \eqref{eq:chiral}.

\subsection{Two-point functions}

The two-point functions of unprotected operators are UV divergent, and consequently they have to be renormalized multiplicatively
\begin{equation}\label{eq:renormalization}
\vec{\cal O} = Z\, \vec{O}
\end{equation}
which generates an anomalous dimension.
In the following I shall use calligraphic ${\cal O}$'s for renormalized operators, with the same index notation as before.
In general, operators can mix under renormalization and consequently have a matrix of anomalous dimensions, as schematically indicated in \eqref{eq:renormalization}.

In order for the three-point functions to possess a conformal structure, the operators must have definite dimension, i.~e.~they have to be eigenstates of the dilation operator and therefore diagonalize \eqref{eq:renormalization}.
Finding such eigenstates is in general a complicated problem, to tackle which the conjectured integrability of the ${\cal N}=4$ SYM spectrum can be exploited.
Note, however, that this does not apply in principle when working at finite $N$.
Nevertheless, in the present situation, the chosen operators and the perturbative order are simple enough that their mixing pattern is almost trivial, which simplifies the computation considerably.
For instance, by conformal symmetry, the twist-2 operators \eqref{eq:twist2} of spin $j$ can mix with all the same spin descendants of twist-2 operators of lower spin $\hat \partial^{j-k}\, O^k$ ($k<j$), giving rise to a mixing pattern governed by a lower triangular anomalous dimension matrix. 
In particular, the twist-2 operators at spin 2 span a space of just two operators, one of which is the descendant of the protected chiral primary $O_{11}$. The operator $\hat O^2$ has no anomalous dimension mixing with the latter at two-loop level.
Further, the two-point functions between twist-2 operators may have non-diagonal finite entries (these are vanishing by construction at tree level thanks to the orthogonality properties of Gegenbauer polynomials, but can arise because of quantum corrections), which can be removed by a finite renormalization, see e.g.~\cite{Belitsky:2007jp}.
This way one obtains an orthogonal set of operators whose two-point functions exhibit the conformal structure
\begin{equation}\label{eq:2point}
\left\langle \hat {\cal O}^j(0) \hat{\bar{ {\cal O}}}^k(x) \right\rangle = C(g^2,N)\, \delta^{jk} \frac{\hat I^j}{|x|^{2\Delta}}
\end{equation}
where $\Delta$ and $j=k$ are the conformal dimension and spin of the operator, $\hat{\bar{ {\cal O}}}$ stands for the conjugate operator, and
\begin{equation}
\hat I \equiv I_{\mu\nu}\, z_1^{\mu}\, z_2^{\nu} \qquad, \qquad I_{\mu\nu} \equiv \eta_{\mu\nu} - 2\, \frac{x_{\mu}x_{\nu}}{x^2}
\end{equation}
with two in principle distinct contractions with null vectors $z_1$ and $z_2$ for the two operators (in practical computations I shall use the same).
In order to normalize the three-point functions suitably, I further re-scale the operators by imposing that the coefficient of \eqref{eq:2point} is $C(g^2,N)=C(g^2,N)^{(0)}$, namely that it coincides with the tree level result and that its quantum corrections are all re-absorbed in the normalization of the operators.

\subsection{Three-point functions}

I consider the following generic three-point functions
\begin{equation}\label{eq:correlator}
\left\langle {\cal O}_{a_1}^{\phantom{a_1}a_2}(x_1)\, \hat {\cal O}^j_{a_3 a_4}(x_2)\, \bar{\cal O}^{a_5 a_6}(x_3) \right\rangle
\end{equation}
From these I extract the following exemplar specific cases:
\begin{subequations}\label{eq:correlators}
\begin{equation*}
\eqref{eq:correlator}\quad\times\quad \left\{ 
\begin{minipage}[c]{0.8\textwidth}

\vspace{-4.5mm}

\begin{alignat}{3}
& \delta_{j0}\, \delta^{a_1}_{1}\, \delta_{a_2}^{2}\, \delta^{a_3}_{2}\, \delta^{a_4}_{3}\, \delta_{a_5}^{3}\, \delta_{a_6}^{1} && \qquad && \text{3 protected}\label{eq:correlators1}\\
& \delta_{j0}\, \delta^{a_1}_{a_2}\, \delta^{a_3}_{1}\, \delta^{a_4}_{2}\, \delta_{a_5}^{2}\, \delta_{a_6}^{1} && \qquad && \text{2 protected, 1 un-protected}\label{eq:correlators2}\\
& \delta_{j2}\, \delta^{a_1}_{2}\, \delta_{a_2}^{1}\, \delta^{a_3}_{1}\, \delta^{a_4}_{1}\, \delta_{a_5}^{1}\, \delta_{a_6}^{2} && \qquad && \text{2 protected, 1 un-protected}\label{eq:correlators3}\\
& \delta_{j2}\, \delta^{a_1}_{a_2}\, \delta^{a_3}_{1}\, \delta^{a_4}_{1}\, \delta_{a_5}^{1}\, \delta_{a_6}^{1} && \qquad && \text{1 protected, 2 un-protected} \label{eq:correlators4}
\end{alignat}
\vspace{-6mm}
\end{minipage}
\right.\nonumber
\end{equation*}
\end{subequations}
Conformal symmetry restricts its functional form to read \cite{Sotkov:1976xe}
\begin{equation}\label{eq:3ptstructure}
\left\langle {\cal O}_{a_1}^{\phantom{a_1}a_2}(x_1)\, \hat {\cal O}^j_{a_3 a_4}(x_2)\, {\cal O}^{a_5 a_6}(x_3) \right\rangle = \frac{{\cal C}_{{\cal O}_{a_1}^{\phantom{a_1}a_2}\, \hat {\cal O}^j_{a_3 a_4}\, {\cal O}^{a_5 a_6}}(g^2,N)\,\hat Y^j}{|x_{12}|^{\Delta_{12,3}-j} |x_{23}|^{\Delta_{23,1}-j} |x_{13}|^{\Delta_{31,2}+j}}
\end{equation}
where I define
\begin{equation}
\hat Y \equiv Y_{\mu}\, z^{\mu}\qquad,\qquad Y^\mu \equiv \frac{x_{12}^\mu}{x_{12}^2} + \frac{x_{23}^\mu}{x_{23}^2}
\end{equation}
and I am using the shorthand notation
\begin{equation}
x_{ij} \equiv x_i - x_j \qquad,\qquad \Delta_{ij,k} \equiv \Delta_i+\Delta_j-\Delta_k
\end{equation}
The dynamics are enclosed in the structure constant ${\cal C}$ which is a function of the coupling $g^2$ and the rank of the gauge group $N$.
The purpose of this note is to compute such coefficients for the correlators \eqref{eq:correlators} in a perturbative expansion at weak coupling up to second order (namely $\lambda^2$, since up to two loops they receive non non-planar corrections).

\section{Strategy}\label{sec:strategy}

The main technical idea behind the computation consists in extracting the structure constant by taking a particular limit of the three-point function.

The limit actually boils down to integrating both sides of \eqref{eq:3ptstructure} over the position of one of the operators, say $x_2$. I am calling this process a limit, because in momentum space such a procedure maps to sending the momentum of the operator sitting at $x_2$ to zero. This method has been applied in \cite{Plefka:2012rd} from which I used various insights\footnote{It has also been applied to the computation of three-point functions in ABJM theory in \cite{Young:2014lka,Young:2014sia}. I redid the computation in \cite{Young:2014lka} and my result is in disagreement with the one quoted there, though.}.
After performing this operation, the perturbative expansion of the left-hand-side of \eqref{eq:3ptstructure} in terms of Feynman diagrams lands on a problem which is technically similar to the computation of an effective two-point function. Consequently, it is much simpler than the original three-point function and can be tackled efficiently, for instance in momentum space. Altogether, the number of diagrams involved in the computation is relatively small (order hundreds) and does not require a tremendous effort. In order to benefit from a consistency test of the computation, I have performed it with an arbitrary gauge fixing parameter (I use a gauge propagator in momentum space proportional to $\frac{1}{k^2}\left(\eta_{\mu\nu}-(1-\alpha)\frac{k_\mu k_\nu}{k^2}\right)$), checking that the dependence on it drops out in the correlation functions.

On the other hand, integrating the right-hand side of \eqref{eq:3ptstructure} produces a bubble integral, with some additional complications due to the tensor structure in the numerator.
After performing such an integration and a Fourier transform, the comparison between the two sides allows to extract the structure constant.

This procedure works efficiently if the integration on the right-hand-side of \eqref{eq:3ptstructure} is finite. 
It may not be the case. Then one could perform the integral in $d$ dimensions (which is the natural choice as the left-hand-side is also computed within dimensional regularization, see remarks below), however the precise functional form of the three-point function in non-integer dimensions may be more complicated and not known, so one would lose predictivity in such cases.
In special circumstances one may still be able to extract sensible information, for instance, if the integration is divergent, but the coefficient of the highest order pole can be mapped unambiguously to the leading order divergence on the other side of \eqref{eq:3ptstructure}, independently of the $\epsilon$ corrections in the functional form of the correlator.
An explicit example arises when integrating in $d=4-2\epsilon$ dimensions over the position of a protected scalar bilinear, with two other scalar operators of generic dimension sitting at the other corners.
In this case, the integration produces a divergence which, if regulated by dimensional regularization, gives a simple pole in $\epsilon$. Its residue maps unambiguously to the structure constant, order by order in the perturbative expansion.
This is cool, as it precisely corresponds to one of the correlators I want to compute, namely \eqref{eq:correlators2}. 

After reduction to a two-point problem, integrals with doubled propagators appear. 
This happens because I am considering only twist-2 operators, for which the inserted operator is connected to two propagators. Sending the momentum of the operator to zero forces the momenta of the propagators to be equal, hence the doubled power.
For the more generic case of composite operators with more fields, the method still applies in principle, however the inserted operator with vanishing momentum acts as a vertex, from the effective two-point function perspective. As a result, the latter gets more and more complicated, with a higher and higher effective loop order.
In other words, one of the shortcomings of the present approach is that it applies most efficiently to low twist operators.

I handle the resulting two-point function integrals applying integration-by-parts (IBP) identities \cite{Chetyrkin:1981qh,Tkachov:1981wb,Laporta:1996mq,Laporta:2001dd} (that I also use for reducing the various numerators arising from the algebra of the diagrams).
To perform this step I have used FIRE5 \cite{Smirnov:2008iw,Smirnov:2013dia,Smirnov:2014hma} and LiteRed \cite{Lee:2012cn,Lee:2013mka}.
This step reduces the expression to a combination of master integrals, which in this a case are known three-loop propagator type \cite{Chetyrkin:1980pr}.
Plugging in their $\epsilon$ expansion, up to the relevant order, I arrive at the final result.

The same three-point function may be integrated with respect to different operator insertion points. If the correlator is not symmetric this provides a non-equivalent computation that can be used as a strong consistency check.

\paragraph{Subtleties with regularization}
I regulate divergences with dimensional regularization.
In order to preserve supersymmetry (and keep a vanishing perturbative $\beta$ function) I use the dimensional reduction scheme \cite{Siegel:1979wq,Avdeev:1980bh,Avdeev:1981vf,Avdeev:1981ew,Velizhanin:2008rw}.
In particular I consider $\delta_a^a = 3 + \epsilon$ complex scalars.
Furthermore, in the definition of the twist-2 operators \eqref{eq:twist2} I use the generalization to $d=4-2\epsilon$ dimensions, which can be read from the expansion of the coefficients of the corresponding Gegenbauer polynomial.
The consequent effects in the renormalization of the operators and their orthogonalization are properly taken into account.

\section{Computation}

\subsection{Two-point functions}\label{sec:2pt}

I first compute the bare two-point functions of all relevant operators, that I need for normalizing the three-point functions and obtain a sensible structure constant.
The two-point functions of the operators of section \ref{sec:operators} can be normalized as in \eqref{eq:2point} with tree-level coefficient, by suitable re-normalizations. In computing them I have retained subleading in $\epsilon$ terms up to the relevant order needed for a consistent two-loop computation in dimensional regularization.
Such expressions are scheme dependent, however in conjunction with the three-point function correlators, they allow to provide scheme independent ratios from which I extract the structure constants in section \ref{sec:structure}.
Therefore I report them for completeness in appendix \ref{app:explicit}.

From their divergent part I extract the anomalous dimensions
\begin{equation}
\gamma_{BPS} = 0 \qquad \gamma_K = \gamma_{\text{spin-2}} = 12 \lambda -48 \lambda ^2 + O(\lambda^3)
\end{equation}
for the protected and the scalar and derivative Konishi operators, respectively. They are in agreement with the known results \cite{Bianchi:2000hn}. 

The spin-2 operators can undergo a mixing with operators of same dimension and spin, in particular the spin-2 descendant of the chiral primary $O_{11}$.
In fact, up to two loops, the matrix of correlators has non-diagonal entries, stemming from finite terms
\begin{equation}\label{eq:finitemix}
\left\langle \hat O^2_{11}(0)\, \hat\partial^2 \bar O^{11}(x) \right\rangle = \lambda^2\left( \frac{1440 \left(N^2-1\right) \hat x^4}{\pi ^4 x^{12}}+O\left(\epsilon\right)\right) + O(\lambda^3)
\end{equation}
These can be eliminated by a finite renormalization of the operator
\begin{align}\label{eq:finitecorr}
\hat {\cal O}^2 = Z_{\text{spin-2}}\, \hat O^2 - \frac{\left\langle \hat O^2_{11}(0)\, \hat \partial^2 \bar O^{11}(x) \right\rangle}{\left\langle \hat \partial^2 O_{11}(0)\, \hat \partial^2 \bar O^{11}(x) \right\rangle}\, \hat \partial^2 O_{11} + O(\lambda^3)
\end{align}
where the correction, according to \eqref{eq:finitemix}, is of order $\lambda^2$.
In the next section I remark the importance of such a correction, in obtaining the correct three-point functions involving these operators.

\subsection{Integrated three-point functions}\label{sec:3pt}

Integrating the structural form of the three-point function on the right-hand-side of \eqref{eq:3ptstructure} over $x_2$ (namely the position of the operator with spin), I find in general
\begin{align}\label{eq:intgen}
\int d^d x_2\, \text{r.h.s.}\eqref{eq:3ptstructure} &= 
\frac{{\cal C}(g^2,N)\, \hat x_{13}^j}{(x_{13}^2)^{\frac{\Delta _1+\Delta _2+\Delta _3+j-d}{2}}}\, 
\frac{\Gamma \left(\Delta _2-\frac{d}{2}\right) \,  
\Gamma \left(\frac{d+j-\Delta _{23,1}}{2}\right) \Gamma \left(\frac{d+j-\Delta _{12,3}}{2}\right)}{\Gamma \left(\frac{\Delta _{12,3}-j}{2}\right) \Gamma \left(\frac{\Delta _{23,1}+j}{2}\right) \Gamma \left(d+j-\Delta _2\right)}\nonumber\\&
\times\, _2F_1\left(
-j, \frac{2-j-\Delta _{23,1}}{2}, \frac{\Delta _{12,3}-j}{2};1\right)
\end{align}
In the present case $\Delta_2 = \theta+j$, $\Delta_1=2+\gamma$, $\Delta_3=2$ and $\gamma=\gamma_K$ for the Konishi or $\gamma=0$ for the protected operator, where $\theta$ is the twist of the operator with spin, including its anomalous correction.

In the special limit $j=0$, the corresponding operator is protected and the relevant correlators are those of the first two lines of \eqref{eq:correlators}. 
Then the expression \eqref{eq:intgen} for the cases at hand simplifies to 
\begin{equation}
\left. \int d^d x_2\, \text{r.h.s.}\eqref{eq:3ptstructure}\right|_{j=0} = {\cal C}(g^2,N)\, \frac{\pi ^{d/2} \Gamma \left(2-\frac{d}{2}\right) \Gamma \left(\frac{d-\gamma -2}{2} \right) \Gamma \left(\frac{d+\gamma -2}{2}\right)}{\Gamma \left(1-\frac{\gamma }{2}\right) \Gamma \left(\frac{\gamma +2}{2}\right) \Gamma (d-2)\, \left(x_{13}^2\right)^{\frac{6+\gamma-d}{2}}}
\end{equation}
Expanding around $d=4-2\epsilon$ and $g=0$ ($g$ appears implicitly in the anomalous dimension), the expression develops a simple pole in $\epsilon$, to all orders in $g$, whose residue is in one-to-one correspondence with the coefficients of ${\cal C}$ in the perturbative expansion.

For the correlators of spin $j>0$ operators, I find 
\begin{align}
\int d^d x_2\, \text{r.h.s.}\eqref{eq:3ptstructure} &= \frac{{\cal C}(g^2,N)\, \hat x_{13}^j}{\left(x_{13}^2\right)^{\frac{\theta +2 j + 4+\gamma-d}{2}}}\, \frac{\pi ^{d/2} \Gamma \left(\frac{d-\gamma-\theta}{2}\right) \Gamma \left(\frac{d+\gamma-\theta}{2}\right) \Gamma \left(\theta+j-\frac{d}{2} \right)}{\Gamma \left(\frac{\gamma+\theta }{2}\right) \Gamma (d-\theta ) \Gamma \left(\frac{2 j-\gamma+\theta}{2}\right)}\nonumber\\&
\times\, _2F_1\left(-j,\frac{1}{2} (-2 j+\gamma-\theta +2);\frac{\gamma+\theta }{2};1\right)
\end{align}
which is instead finite at $d=4$. The coefficients of the expansion in $g$ mix those of the structure constant ${\cal C}(g^2,N)$ and of the anomalous dimensions, nevertheless one can invert \eqref{eq:3ptstructure} and fix the latter. 
I remark that integrating over $x_2$ corresponds to sending the momentum of the operator with spin to 0. This suppresses the mixing with descendants, which drops from the computation.

The expression \eqref{eq:3ptstructure} can also be integrated over the insertion point of a scalar operator, namely $x_3$. In that situation 
\begin{align}
\int d^d x_3\, \text{r.h.s.}\eqref{eq:3ptstructure} &= 
\frac{{\cal C}(g^2,N)\, \hat x_{12}^j}{\left(x_{12}^2\right)^{\frac{\theta +2 j + 4+\gamma-d}{2}}}\, \frac{\pi ^{d/2} \Gamma \left(2-\frac{d}{2}\right) \Gamma \left(\frac{d+\gamma-\theta}{2}\right) \Gamma \left(\frac{\theta+d-4-\gamma}{2} \right)}{ \Gamma (d-2 )\Gamma \left(\frac{4+\gamma-\theta }{2}\right) \Gamma \left(\frac{\theta-\gamma}{2}\right)}\nonumber\\&
\times\, _2F_1\left(2-\frac{d}{2},-j;\frac{\theta-\gamma}{2};1\right)
\end{align}
The expansion of the latter in $\epsilon$ is again divergent, with a simple pole at each perturbative order.
In this case the mixing with descendants is not negligible any longer.

When integrating these three-point functions over the insertion point of a protected operator, as that located at $x_3$, the result is divergent, with a simple pole in the regulator $\epsilon$, whose residue is proportional to the structure constant. One can thus prefer such an integration, since the required order for the $\epsilon$ expansion of the diagrams (and therefore of the integrals) is lower and therefore the computation a bit more economic.
It would be even more efficient to use this piece of information to restrict the number of diagrams from the onset, on the basis of their divergence properties. This yields generally powerful simplifications, for instance when computing the anomalous dimensions of operators, where one disregards diagrams that are not UV divergent by power counting. This is especially strong when working in superspace formalism, where supergraphs have improved UV properties. However, in the case at hand, the poles in $\epsilon$ emerge from a mixture of UV and IR effects, the latter produced by the soft limit of vanishing momentum, imposed on one of the operators in the correlators.
Therefore, the UV arguments mentioned above do not apply straightforwardly to this case.
Moreover, a complete computation requires not only the knowledge of the bare structure constants, but also of the two-point functions of the operators. For these, the relevant diagrams and integral expansions are structurally the same as for the three-point functions (apart from combinatorics and different powers of the propagators), and have to be carried out up to finite order in $\epsilon$ anyway. This is just to remark that I do not see any striking advantage of integrating over a particular operator (provided the integration is not badly divergent as stressed above), and that the different integrations entail computationally equivalent calculations. 

\section{Structure constants}\label{sec:structure}

Using the results of the previous section I am finally able to compute the structure constants of various three-point functions, involving the operators defined in section \ref{sec:operators}.
I recall that the operators are renormalized in such a way that their two-point function coincides with the tree-level one.
Moreover I take the ratio between the structure constants and their tree-level expressions.
The final results are the scheme independent weak coupling expressions of the structure constants of the correlators \eqref{eq:correlators}, whose perturbative expansions read
\begin{alignat}{2}
& \left(\frac{{\cal C}}{{\cal C}^{(0)}}\right)_{\eqref{eq:correlators1}}\hspace{-0.9cm}(\lambda)\quad = 1 +O\left(\lambda ^3\right)  && \text{BPS$^3$}\label{eq:CPO}\\
& \left(\frac{{\cal C}}{{\cal C}^{(0)}}\right)_{\eqref{eq:correlators2}} \hspace{-0.9cm}(\lambda)\quad = 1-6 \lambda +\lambda ^2 (36 \zeta (3)+66)+O\left(\lambda ^3\right)  && \text{Konishi, BPS$^2$}\label{eq:3ptKon}\\
& \left(\frac{{\cal C}}{{\cal C}^{(0)}}\right)_{\eqref{eq:correlators3}} \hspace{-0.9cm}(\lambda)\quad = 1-6 \lambda +\lambda ^2 (36 \zeta (3)+66)+O\left(\lambda ^3\right) \qquad\qquad  && \text{spin-2, BPS$^2$}\label{eq:3ptCPOtwist}\\
& \left(\frac{{\cal C}}{{\cal C}^{(0)}}\right)_{\eqref{eq:correlators4}} \hspace{-0.9cm}(\lambda)\quad = 1-3 \lambda +21 \lambda ^2+O\left(\lambda ^3\right)  && \text{Konishi, spin-2, BPS}\label{eq:3ptKontwist}
\end{alignat}
The three-point function of three protected operators \eqref{eq:CPO} does not receive quantum corrections, as expected from non-renormalization theorems \cite{Lee:1998bxa,Eden:1999gh,Arutyunov:2001qw,Heslop:2001gp}.
The two-loop structure constant for one scalar Konishi in the singlet of $SU(4)$ and two protected operators \eqref{eq:3ptKon} coincides with the result computed via the OPE expansion of four-point correlators of BPS operators, and constitutes an independent test of that. This also coincides with the structure constant of the derivative operator with spin 2 and two protected operators \eqref{eq:3ptCPOtwist}.
The last correlator \eqref{eq:3ptKontwist} of one protected and two un-protected operators is a (to the best of my knowledge) novel prediction and constitutes the main result of this note. 
It would be interesting to derive it also from an OPE expansion (where also higher spin operators would be more easily accessible), for instance building on the four-point correlators computed in \cite{Bianchi:2002rw}.

As a consistency check of my result, I re-computed the same structure constants, by integrating on a different insertion point, namely over $x_3$, at the position of the protected operator (I recall that integrating over the position of the scalar Konishi produces divergences whose order in $\epsilon$ increases with the number of loops, spoiling the comparison with the Feynamn diagram computation). 
In Fourier space, the momentum of the spin-2 operator is no longer vanishing as in the previous case, which means that the correlator is sensitive to a potential finite mixing with the descendant of the chiral operator $ $.
Indeed, naively computing the integrated correlator using the operator $\hat O^2$, one gets incorrect results, different from \eqref{eq:3ptCPOtwist} and \eqref{eq:3ptKontwist}.
Including the finite correction \eqref{eq:finitecorr}, which amounts to adding a contribution proportional to the tree level schematic correlators $\left\langle O_{BPS} \hat \partial^2 O_{BPS} \bar O_{BPS} \right\rangle$ and $\left\langle O_K \hat \partial^2 O_{BPS} \bar O_{BPS} \right\rangle$, respectively, precisely compensates for such  a mismatch and grants reproducing \eqref{eq:3ptCPOtwist} and \eqref{eq:3ptKontwist}. Here I have renamed the operators more simply according to their properties as ${\cal O}_{BPS}$ for protected operators and their conjugates and ${\cal O}_{K}$ for the scalar Konishi.

Finally, from the latter computation, I can also provide the structure constant of the scalar Konishi with a chiral primary and a spin-2 descendant (this is not available integrating over the position of the latter operator, $x_2$, since it produces a vanishing result as a consequence of the zero momentum limit).
The result reads
\begin{equation}
\frac{{\cal C}_{\text{K,$\partial^2$BPS,BPS}}}{{\cal C}^{(0)}_{\text{K,$\partial^2$BPS,BPS}}} = 1+3 \lambda +\lambda ^2 (36 \zeta (3)-6)+O\left(\lambda ^3\right)
\end{equation}
Replacing the scalar Konishi with a protected operator I find again a trivial result, as expected.
It would be interesting to extend this analysis to operators with higher spin and perhaps find expressions for generic spin, multiple operators with spin and also provide a derivation from the OPE of suitable four-point functions and within the integrability approach.

\acknowledgments

This work has been supported by DFF-FNU through grant number DFF-4002-00037.

\vfill
\newpage

\appendix

\section{Explicit expressions}\label{app:explicit}

In this appendix I collect the explicit expressions of the re-normalizations of the operators and the integrated three-point functions computed by evaluating Feynman diagrams.

The expressions of the two-point function re-normalizations read (here I have also collected factors to get rid of logarithms in the correlators, for convenience)
\begin{align}
Z_{BPS} &= 1+ \lambda  \left(12\zeta (3) \epsilon + O(\epsilon^2)\right) + \lambda^2\, O(\epsilon)\nonumber\\
Z_{K} &= 1+\lambda \pi ^{\epsilon } x^{2 \epsilon } \left(\frac{6}{\epsilon }+8+\frac{1}{2} \left(24 \zeta (3)+\pi ^2+4\right) \epsilon + O(\epsilon^2) \right)
\nonumber\\&+\lambda ^2 \pi ^{2 \epsilon } x^{4 \epsilon } \left(\frac{18}{\epsilon ^2}+\frac{36}{\epsilon}+36 \zeta (3)+3 \pi ^2-14 + O(\epsilon)\right)+O\left(\lambda ^3\right)\nonumber\\
Z_{\text{spin-2}} &= 1+\lambda  \pi ^{\epsilon } x^{2 \epsilon } \left(\frac{6}{ \epsilon }-1 + \frac{1}{2} \left(24 \zeta (3)+ \pi ^2-14\right) \epsilon + O(\epsilon^2)\right)
\nonumber\\&
+\lambda ^2 \pi ^{2 \epsilon } x^{4 \epsilon } \left(\frac{18}{\epsilon ^2} -\frac{18}{\epsilon} + 36 \zeta (3) +3 \pi ^2 -41 + O(\epsilon)\right)+O\left(\lambda ^3\right)
\end{align}
From the divergent part I extracted the anomalous dimensions of section \ref{sec:2pt}, evaluating $-\displaystyle\lim_{\epsilon\to 0}\, \mu\frac{d \log Z}{d\mu}$.
The subleading in $\epsilon$ terms, up to the corresponding relevant order for each loop, are needed for consistency with dimensional regularization at two loops.
Especially, the protected operators have also a protected two-point function \cite{Penati:1999ba}, but here I have also retained subleading $O(\epsilon)$ terms.

From the perturbative computation I find the following integrated three-point functions
\begin{align}
\int d^d x_2\, \eqref{eq:correlators1} &= -\frac{N^2-1}{64 \epsilon\, \pi ^{4-2 \epsilon} x_{13}^{2-4 \epsilon} } \left(\left(1+\frac{\pi ^2 \epsilon ^2}{6}+O(\epsilon^3)\right) + \lambda  \left(-12\zeta (3) \epsilon+O(\epsilon^2)\right)
 \right. \nonumber\\&\left.
+ \lambda^2\, O(\epsilon)
\right) + O(\lambda^3)
\end{align}
\begin{align}
\int d^d x_2\, \eqref{eq:correlators2} &= -\frac{N^2-1}{32 \epsilon\, \pi ^{4-2 \epsilon} x_{13}^{2-4 \epsilon} }
\left(\left(1 +\frac{\pi ^2 \epsilon ^2}{6} + O(\epsilon^3)\right)
 \right. \nonumber\\&
 - \lambda  \pi ^{\epsilon } x_{13}^{2 \epsilon } \left(\frac{6}{\epsilon }+14 + \frac{1}{2} \left(24 \zeta (3)+3 \pi ^2+56\right) \epsilon + O(\epsilon^2)\right) \nonumber\\&\left. +
\lambda ^2 \pi ^{2 \epsilon } x_{13}^{4 \epsilon } \left( \frac{18}{\epsilon ^2}+\frac{96}{\epsilon }+6 \left(24 \zeta (3)+\pi ^2+62\right) + O(\epsilon)\right)\right) + O(\lambda^3)
\end{align}
\begin{align}
\int d^d x_2\, \eqref{eq:correlators3} &= \frac{3 \left(N^2-1\right) \hat x_{13}^2}{4\, \pi ^{4-2 \epsilon} x_{13}^{6-4 \epsilon}} 
\left( \left(1-\frac{11 \epsilon }{3}+\left(4+\frac{\pi ^2}{6}\right) \epsilon ^2 + O(\epsilon^3)\right)
\right.\nonumber\\&
-\lambda  \pi ^{\epsilon } x_{13}^{2 \epsilon } \left(+\frac{6}{\epsilon }-30+\frac{1}{6} \left(192 \zeta (3)+9 \pi ^2+316\right) \epsilon + O(\epsilon^2)\right)
\nonumber\\&\left.
+ \lambda ^2 \pi ^{2 \epsilon } x_{13}^{4 \epsilon } \left(\frac{18}{\epsilon ^2}-\frac{102}{\epsilon }+ 264 \zeta (3)+6 \pi ^2+260 + O(\epsilon)\right)
\right) + O(\lambda^3)
\end{align}
\begin{align}
\int d^d x_2\, \eqref{eq:correlators4} &= \frac{3\left(N^2-1\right) \hat x_{13}^2}{2\, \pi ^{4-2 \epsilon} x_{13}^{6-4 \epsilon}}
\left( \left(1-\frac{11 \epsilon }{3}+\left(4+\frac{\pi ^2}{6}\right) \epsilon ^2 + O(\epsilon^3)\right)
\right.\nonumber\\&
-\lambda \pi ^{\epsilon } x_{13}^{2 \epsilon } \left(\frac{12}{\epsilon }-47+\frac{1}{3} \left(96 \zeta (3)+9 \pi ^2+160\right) \epsilon + O(\epsilon^2)\right) 
\nonumber\\&\left.
+ \lambda ^2 \pi ^{2 \epsilon } x_{13}^{4 \epsilon } \left(\frac{72}{\epsilon ^2}-\frac{276}{\epsilon }+8 \left(57 \zeta (3)+3 \pi ^2+46\right) + O(\epsilon)\right)
\right) + O(\lambda^3)
\end{align}
Taking the ratios with the proper normalizations of the operators and comparing these to the expected expressions of section \ref{sec:3pt}, I derived the structure constants of section \ref{sec:structure}.

\bibliographystyle{JHEP}

\bibliography{biblio2}

\end{document}